\documentclass[a4paper,11pt]{article}
\usepackage{epsfig}
\usepackage{amssymb}
\begin{document}
\thispagestyle{empty}

\newcommand{\etal}  {{\it{et al.}}}  
\def\Journal#1#2#3#4{{#1} {\bf #2}, #3 (#4)}
\def\PRD{Phys.\ Rev.\ D}
\def\NIMA{Nucl.\ Instrum.\ Methods A}
\def\PRL{Phys.\ Rev.\ Lett.\ }
\def\PLB{Phys.\ Lett.\ B}
\def\EPJ{Eur.\ Phys.\ J}
\def\IEEETNS{IEEE Trans.\ Nucl.\ Sci.\ }
\def\CPCD{Comput.\ Phys.\ Commun.\ }

\smallskip

\bigskip

{\huge
\begin{center}
Gauge model of unparticles
\end{center}
}


\begin{center}
{\Large
 G.A. Kozlov

 }
\end{center}


\begin{center}
\noindent
 {\large
 Bogolyubov Laboratory of Theoretical Physics\\
 Joint Institute for Nuclear Research,\\
 Joliot Curie st., 6, Dubna, Moscow region, 141980 Russia\\

 }
\end{center}


\begin{abstract}

\noindent Modern theoretical models strongly suggest that new
phenomena await discovery above the energy scale of the Standard
Model (SM) of particle interactions. In this paper we argue that
correct description of particle physics in the TeV energy scale
needs to account for degrees of freedom obeying the conformal (or
scale) invariance. In this respect, the existing of Georgi's
unparticles is strongly argued. We present the gauge model of
scalar unparticles. The ground state of scalar unparticles in the
continuum is restricted by the vacuum expectation value of the SM
Higgs boson and the scale dimension $d$. In the framework of
Abelian gauge field theories we develop a model in
four-dimensional space-time in which the scalar unHiggs is a
dipole field. The quantization is performed within the canonical
formalism. Based on general principles of quantum field theory the
new features of unHiggs propagators are carried out for the first
time.
\end{abstract}

PACS 11.10.Cd, 11.15.Ex
\section{Introduction}

\bigskip

The unparticle phenomenon [1] and several points concerning unparticle
phenomenology have been
widely overlooked in the
literature (see, e.g., the recent papers in [2]).
However,
to the author's opinion, a set of matter things has not been
settled in a way suitable to more better understanding of the
subject. This paper is an attempt to clarify the unparticle nature
in gauge theories, e.g., within the invariance under gauge
transformation of the second kind. Since unparticles can couple as
a Standard Model (SM)-like singlet or even doublet to the SM
particles, they must carry the SM - like charges. A gauge model of
unparticles must also contain the criteria relevant to SM pattern,
however to keeping in mind non-canonical scaling dimension of the
unparticle.

In general, a conformal theory, is one where there is an exact
scale invariance (apart from more technical aspects): conformal
invariance implies scale invariance, theory looks the same on all
scales. SM is not conformal even as a classical field theory - the
mass of the Higgs boson breaks conformal symmetry. One of the
basic features of conformal theory is there are no masses. One can
really deal with the theory so that predictions are possible in
case this theory is renormalizable. The Higgs field is necessary
to guarantee the renormalizability. On the other hand, at high
energies, the Higgs field can weakly interact with a hidden
scale-invariant sector realized through the unparticles staff
which would be relevant since we do hope to observe the Higgs
particle in experiment. The Higgs field  can serve as a portal to
a hidden sector of the SM matter. Such interactions can be
systematically described in the framework of the field theory with
the effective Lagrangian density (LD).

We suppose the effective field theory (EFT) containing a hidden
sector lying beyond the SM. We model this hidden sector using an
arbitrary field operator $O(M)$ on the running mass scale $M$ in
interaction with a heavy state that may occur in the ultraviolet
(UV) region at high scale $\Lambda
>> \Lambda_{SM}$, where $\Lambda_{SM} \sim O (v)$  with
$v\simeq 246$ GeV being the vacuum expectation value of the Higgs
field in the SM. The simplest form of the relevant LD is
\begin{equation}
\label{e01} L_{EFT}(M) = c(M,\Lambda)\, O(M) = \frac{c_{0}
(M)}{\Lambda^{d-4}}\, O(M),
\end{equation}
where $d$ is the scaling dimension of $O(M)$ and the effect of
heavy state is encoded in $c(M,\Lambda)$. Physically interesting
goal is to develop the theory near its infra-red (IR) fixed point
$M_{IR}\simeq\Lambda_{SM}$.

Somewhat surprisingly it is possible to have a model that has
basically the singlet-doublet mixing. One can start with an
interaction of the form $\sim O_{U}\,H\,H^{+} +
O_{U}^{2}\,H\,H^{+}$, where $O_{U}$ is the new singlet-like
unparticle field, or even the unHiggs staff (a part of the
approximate conformal field theory) with $1 < d < 2$, and $H$
stands for the SM Higgs field with $Re H = (h+v)/\sqrt{2}$. Both
$H$ and $O_{U}$ fields receive their vacuum expectation values and
one ends up with the Higgs particle and unHiggs staff,
respectively.

When the staff $O_{U}$ is added to the Higgs scalar, two effects
can appear: the mixing and the invisible decay. In case that there
is an unbroken symmetry in the singlet sector (of unparticles) the
singlets are stable and weakly interacting.

We find that the behaviour of the unHiggs field is compatible with
that of a dipole field. In author's opinion this fact looks both
attractive and instructive. In this spirit the model is looked as
a first attempt to start with an effective Lagrangian at energies
$E$ covering the region between the IR scale and the
UV scale, $\tilde\Lambda_{U} < E <  \Lambda_{U}$. To our
knowledge, independently of the validity of the conjecture of
references (see, e.g., the references in [2]), a discussion
concerning the unparticle states of dipole field models with the
dimension $d$ is lacking in current literature. In this paper we
give a positive contribution to this direction by making use the
standard Lagrangian approach to quantum field theory, in
particular, the canonical formalism and symmetry properties.

What can one say about the collider signals of unparticles. There
are, at least, two possibilities:\\
 - virtual unparticle exchanges \
 and\\
 - real production of unparticles, e.g., in gluon-gluon fusion,
 $gg\rightarrow gU$.\\
 If $U$-unparticle does not decay one gets monojets. More
 generally, we would get missing energy signals. On the other
 hand, unparticle can decay to the SM particles. This will
 severely modify the set of signals considered above. We do not
 get missing energy signals in general.

Since unparticles are light enough (formally, due to scale
invariance the spectrum of them is continuous) the Higgs boson in
wide mass region can decay into these invisible unparticles,
thereby leaving no obvious signal at the LHC or ILC and muon
collider. We have no a task to investigate this issue, and the
references can be found elsewhere.

The outline of the paper is as follows. We will review the
inclusion of  interactions $\sim O_{U}\,H\,H^{+} +
O_{U}^{2}\,H\,H^{+}$ for an unparticle field in the next section.
In section 3 we will examine the gauge unHiggs model where this
unparticle is a dipole field. The quantization of the "unHiggs"
field is given in the section 4. In section 5 we give the
propagator of the unHiggs field. Our conclusion is presented in
the last section.

\section{An extended Higgs - unparticle tower model}
\label{bec}

We start with the effective LD

\begin{equation}
\label{e1} L = L_{1} + L_{2},
\end{equation}
where
\begin{equation}
\label{e2} L_{1} = -\frac{1}{4}F_{\mu\nu} F^{\mu\nu} +
\alpha^{2}\,D_{\mu} O_{U}\,\bar D^{\mu} O^{\ast}_{U} -
\lambda^{2}\,{\vert O_{U}\vert}^{4} + \mu^{2}_{0}\,{\vert
O_{U}\vert}^{2},
 \end{equation}
\begin{equation}
\label{e3} L_{2} = a\,{\vert H\vert}^{2}\,O_{U} + b\,{\vert
H\vert}^{2}\,{\vert O_{U}\vert}^{2}.
\end{equation}
In formulae (\ref{e2}) and (\ref{e3}),  $F_{\mu\nu} =
\partial_{\mu}A_{\nu} -
\partial_{\nu}A_{\mu}$ with the gauge field $A_{\mu}$;
$D_{\mu} =\partial_{\mu} + ieA_{\mu}$, $\bar D_{\mu}
=\partial_{\mu} - ieA_{\mu}$; $\alpha$, $\lambda$, $\mu_{0}$, $a$
and $b$ are $d$-dependent in mass units: $1-d$, $2(1-d)$, $2-d$,
$2-d$ and $2(1-d)$, respectively.

The scaling properties of hidden sector depend on the scaling
properties of couplings. There are three characteristic scales
$\Lambda$ at high energies in the theory.\\
1. $M_{U}$ scale: hidden sector (unparticles) is coupled to SM
through non-renormalizable couplings at $M_{U}$. This leads to the
interaction form $M_{U}^{-k}\,O_{SM}\,O_{BZ}$, where $O_{SM}$ is
the SM operator, while $O_{BZ}$ stands for Banks-Zaks theory [3]
operator (massless fermions case). Here, UV sector coupled to SM
through the exchange
of messenger fields with large mass scale $M_{U}$.\\
2. $\Lambda_{U}$ scale: assume that unparticle sector becomes
conformal at $\Lambda_{U} < M_{U}$, and couplings to SM preserve
conformality in IR region that gives the operator form $
M_{U}^{-k}\,O_{SM}\,O_{BZ}\rightarrow const\,\Lambda ^{-d-n+4}
O_{SM}^{n}\,O_{U}$. Here, the messenger sector decouples resulting
in contact interaction between SM and an unparticle sector. The
latter flows into a non-perturbative IR fixed point hence
exhibiting scale invariance.\\
3. Electroweak symmetry breaking leads to conformal symmetry
breaking through the super-renormalizable operator $a\,{\vert
H\vert}^{2}\,O_{U}$ at the third scale $\tilde\Lambda_{U} =
(a\,v^{2})^{1/(4-d)} < \Lambda_{U}$. Thus when the Higgs field
accepts its vacuum expectation value (v.e.v.) the operator
$a{\vert H\vert}^{2} O_{U}$ brings a scale (occurrence of a
tadpole term in $L_{2}$). A sector of unparticle staff leaves the
conformal-fixed point and a theory becomes nonconformal-invariant
at the scale $\tilde\Lambda_{U}$ below which the unparticle staff
transforms into a standard particle sector.

One can conclude that unparticle physics is only possible in the
conformal window. However, the width of this window depends on
$d$, $\tilde\Lambda_{U}$, $\Lambda_{U}$, $M$.

 Actually, the LD (\ref{e2}) is obtained by switching on a
gauge field ($A_{\mu}$) interaction in the pure $O_{U}$-field LD
\begin{equation}
\label{e22} L_{0} = \alpha^{2}\partial_{\mu} O_{U}\,\partial^{\mu} O_{U}-
\lambda^{2}\,{\vert O_{U}\vert}^{4} + \mu^{2}_{0}\,{\vert
O_{U}\vert}^{2}.
 \end{equation}

In terms of the energy scale $E\sim \Lambda$ we define $\alpha =
\alpha (\Lambda)$ and $\alpha (\Lambda)\,O_{U} (x) \rightarrow
\phi (x)$ at $\Lambda < \tilde \Lambda _{U}$, where $\phi (x)$ is
the SM scalar field. A crucial role is the weak coupling of the
unparticle fields to the SM ones, in particular to Higgs boson
$H$.
Ignoring these couplings will lead to unability to "observe"
unparticle staff. Although the precise investigation of gauge
higgsless LD (\ref{e2}) defines the dipole behaviour of the
unHiggs field (see sections 3, 4 and 5).

The nature of unparticle staff is unknown and one can only use one
of the schemes relevant to the concrete task. To find the
nontrivial result, in particular, to get the v.e.v. of unparticle
staff, one should follow a couple of steps:

1. to establish the structure of $O_{U}$ operator: we choose the
scheme within the infinite tower of $N$ - scalars $\phi_{k}$
$(k=1, 2, ..., \infty$) [4]
\begin{equation}
\label{e4} O_{U}\rightarrow O = \sum^{N=\infty }_{k=1}
f_{k}\,\phi_{k},
\end{equation}
where the scalar fields $\phi_{k}$ are characterized by the mass
squared $m^{2}_{k} = k\,\Delta^{2}$ as $\Delta\rightarrow 0$; \\
2. the LD (\ref{e1}) is presented in the form:
\begin{equation}
\label{e5} L = -\frac{1}{4}F_{\mu\nu} F^{\mu\nu} +
\alpha^{2}\,D_{\mu} O\,\bar D^{\mu} O^{\ast} -
 V(O,H),
\end{equation}
where the potential $V(O,H)$ is given in the deconstructed
conventional form
\begin{equation}
\label{e6} V(O,H) \rightarrow  V(\phi_{k},H) = \frac{1}{4}\,\xi
\left (\sum^{N}_{k=1} \phi^{2}_{k}\right )^{2} -
\frac{1}{2}\sum^{N}_{k=1} m^{2}_{k}\,\phi^{2}_{k} - a\,{\vert
H\vert}^{2}\,\sum^{N}_{k=1} f_{k}\,\phi_{k}-
 b^{\prime} {\vert H\vert}^{2}\,\sum^{N}_{k=1} \phi^{2}_{k}
\end{equation}
with new coefficients $\xi$ and $b^{\prime}$ in zero mass units.
Obviously, the LD (\ref{e5}) has the same pattern as those in
(\ref{e1}) within the deconstructed convention. The minimization
equation for $\phi_{k}$-fields looks like at
$\langle\phi_{k}\rangle = \sigma_{k}$ and $\langle H\rangle =
v/\sqrt{2}$
\begin{equation}
\label{e7} \sigma_{k} = \frac{1}{2}\,\frac{-
a\,v^{2}\,f_{k}}{m^{2}_{k}-  (\xi \sum^{N}_{l=1} \sigma^{2}_{l}  -
 b^{\prime}\,v^{2})}.
\end{equation}
It is evidently from (\ref{e7}) that the interaction term
$a\,{\vert H\vert}^{2}\,O_{U}$ in (\ref{e3}) with $a\neq 0$
ensures $\sigma_{k}\neq 0$.

 The free propagator of the field $O_{U}$ is
\begin{equation}
\label{e8} D(p^{2},d)
=\int_{0}^{\infty}\frac{dm^{2}}{2\,\pi}\,\frac{\rho_{0}(m^{2},d)}
{p^{2}- m^{2} + i\,\epsilon},
\end{equation}
where in accordance with the scale invariance the spectral density
$\rho_{0}(m^{2},d)$ is
\begin{equation}
\label{e9} \rho_{0}(m^{2},d) = A_{d}\,{\left (m^{2}\right)}^{d-2}.
\end{equation}
The $A_{d}$ is given by the condition $\rho_{0}(m^{2},d=1) =
m^{-2}$ and has the form [1]
\begin{equation}
\label{e10} A_{d} =
\frac{16\,\pi^{5/2}}{(2\,\pi)^{2d}}\,\frac{\Gamma (d+1/2)} {\Gamma
(d-1)\,\Gamma(2d)}.
\end{equation}
On the other hand, $\rho_{0}(m^{2})$ is expressed in the form:
\begin{equation}
\label{e11} \rho_{0}(m^{2})= 2\,\pi\sum_{\omega}\delta (m^{2} -
m^{2}_{\omega})\,{\vert\langle 0\vert
O(0)\vert\omega\rangle\vert}^{2},
\end{equation}
where the sum is over all relativistic states $\vert\omega\rangle$
normalized by relativistic manner.
From (\ref{e9}) and (\ref{e11}) one finds
 (${\vert\langle 0\vert O(0)\vert\omega_{k}\rangle\vert}^{2} = f^{2}_{k}$):
\begin{equation}
\label{e12} f^{2}_{k} = \frac{A_{d}}{2\,\pi}\,\left
(m^{2}_{k}\right )^{d-2}\,\Delta^{2}.
\end{equation}
In the continuum
\begin{equation}
\label{e13}
 \langle O_{U}\rangle = \int^{\infty}_{0}
\bar\sigma(s)\bar f(s)ds =
\frac{v^{2}}{2}\int^{\infty}_{0}\frac{\bar f^{2}(s)}{z-s}ds=
\frac{A_{d}}{4\pi}av^{2}(-1)^{d-1}z^{d-2}\Gamma (d-1)\Gamma (2-d),
\end{equation}
where $ z = b^{\prime}\,v^{2} - \xi \sum^{N}_{l=1} \sigma^{2}_{l}$
is the infra-red regularized (IRR) mass square (mass gap),
induced by interaction ${\vert H\vert}^{2}O_{U}$
in the unparticle continuum.
This mass vanishes when the ratio
between $v$ and $\sigma$ is given by $\xi/b^{\prime}$.

We find that the v.e.v. of the $O$ - field is defined only for
nonzero $v^{2}$. In case of weak interaction ($a\rightarrow 0$)
between the $O$ - field and the Higgs-field, $\sigma\rightarrow
0$. As the result, $\sigma^{2}\neq 0$ only if  $v^{2}\neq 0$ so
that the IRR - mass is provided by the electro-weak symmetry
breaking, and this mass does cutoff the infra-red divergence of
the $O$ - field for $b^{\prime}$ and $\xi$ - couplings obeying the
condition
\begin{equation}
\label{e16} \frac{b^{\prime}}{\xi} > \frac{\sigma^{2}}{v^{2}}
\end{equation}
which leads to new understanding how to avoid the infra-red
trouble.

\section{ The gauge "unHiggs" model}
\label{bec} The gauge "unHiggs" model is naturally obtained itself
by considering the LD (\ref{e2}) which is invariant under the
transformations
\begin{equation}
\label{e17} A_{\mu}(x)\rightarrow A_{\mu}(x) +
\partial_{\mu}\vartheta (x),
\end{equation}
\begin{equation}
\label{e18} O_{U}(x)\rightarrow\exp [-i\,e\,\vartheta
(x)]\,O_{U}(x),
\end{equation}
where $\vartheta (x)$ is a real function obeying the equation
$\nabla^{2}\vartheta (x) = 0$ ($\nabla \equiv
\partial/\partial\,x_{\mu}$).

The generating current \begin{equation} \label{e19} K_{\mu}
(x;\Lambda) = i\,\alpha^{2}(\Lambda) \left
[O^{\ast}_{U}(x)\,D_{\mu}\,O_{U}(x) - O_{U}(x)\,\bar
D_{\mu}\,O^{\ast}_{U}(x)\right ]
\end{equation}
enters in the equation of motion in the standard form
\begin{equation}\label{e20} K_{\mu} (x) = \partial^{\nu}F_{\nu\mu}
(x) = \nabla^{2} A_{\mu}(x) - \partial^{\nu}\partial_{\mu} A_{\nu}
(x).
\end{equation}

Because of the physical reason and in order to be more close to
the standard field theory one can write the current (\ref{e19}) in
terms of the field operators occurred in the limiting procedure.
For simplicity let us introduce the scalar field $\Phi (x;\nu)$
where $x$ is a Minkowski coordinate while the fifth coordinate
$\nu$ is thought as a continuous index. The operator $O_{U}(x)$
can then be defined in terms of the field $\Phi (x;\nu)$ as
follows:
\begin{equation}
\label{e201} O_{U}(x) = \lim_{{\nu}\rightarrow 0} \nu^{- d}\, \Phi
(x;\nu)
\end{equation}
and hence
\begin{equation}
\label{e202} K_{\mu} (x;\Lambda) = i\alpha^{2}
(\Lambda)\,\lim_{\nu,\nu^{\prime} \rightarrow 0}\nu^{-
d}\,(\nu^{\prime})^{-d}\left [\Phi^{\ast} (x;\nu)\,D_{\mu}\,
\Phi(x;\nu^{\prime}) - \Phi (x;\nu^{\prime})\,\bar D_{\mu}\,\Phi
^{\ast} (x;\nu)\right ].
\end{equation}

The SM operators couple to the field $\Phi (x;\nu)$ only on the
boundary $\nu\rightarrow 0$. Equation (\ref{e202}) can not be
tenable and satisfactory obey the Equation (\ref{e20}) at $\Lambda
> \tilde\Lambda_{U}$ where, again, $ \tilde\Lambda_{U}$
corresponds to the scale below which the SM world interactions
occurred. The unparticle phase has no conventional particle
description. The only remarkable characteristic of unparticles is
that the phase space in decay processes to unparticle staff is the
same as the phase space for decay to $d$ massless particles. To
avoid this dissatisfaction we introduce the hidden parameter
$\beta_{U}$ in the kinetic term for $A_{\mu}(x)$ field in
(\ref{e2}) leading to modified Equation (\ref{e20})
\begin{equation}
\label{e203} K_{\mu} (x;\Lambda) = \nabla^{2} A_{\mu}(x) -
\beta_{U}\,\partial^{\nu}\partial_{\mu} A_{\nu} (x), \,\,\,
\Lambda > \tilde\Lambda_{U}, \,\, \beta_{U} \neq 1.
\end{equation}
It is closely related to a model introduced by Ferrari [6] in a study of
the Higgs phenomenon, from which the present model is distinguished by
coupling to unHiggs staff.
Obviously, $\beta_{U} = 1$ if $\Lambda < \tilde\Lambda_{U}$. The
stabilization mechanism (\ref{e203}) is also valid if one uses the
infinite tower scheme for $O_{U}(x)$ operator (\ref{e4}). It is closely

 To find a solution of the "unHiggs" model we introduce the
real fields $\psi$ and $\chi$ in
\begin{equation}
\label{e21} O_{U} = \frac{1}{\sqrt{2}} (\sigma + \psi +
i\,\chi),\,\,\, O^{\ast}_{U} = \frac{1}{\sqrt{2}} (\sigma + \psi -
i\,\chi),
\end{equation}
where $(\Omega, \chi\Omega) = 0$, $(\Omega, (\sigma + \psi)\Omega)
= \sigma \neq 0$ and $\Omega$ is the vacuum. In terms of new real
fields $\psi$ and $\chi$ the Lagrangian density (\ref{e2}) of the
order $e\sigma$ and $\lambda\sigma$ becomes
\begin{eqnarray}
\label{e22} \tilde L_{1} =
-\frac{1}{2}\,(\partial_{\mu}A_{\nu})^{2} +
\frac{1}{2}\beta_{U}\,\,\partial_{\mu}A_{\nu}\partial^{\nu}A^{\mu}-
\frac{1}{2}\,m^{2}A^{2}_{\mu} + \frac{1}{2}\, \alpha^{2}
[(\partial_{\mu}\chi)^{2} + (\partial_{\mu}\psi)^{2}]\cr +
m\,\alpha A^{\mu}\,\partial _{\mu}\chi - \frac{1}{2}\,
(3\,\lambda^{2}\,\sigma^{2} - \mu^{2}_{0})\,\psi^{2} -
\frac{1}{2}\, (\lambda^{2}\,\sigma^{2} - \mu^{2}_{0})\chi^{2},
\end{eqnarray}
where $m = e\,\alpha\,\sigma$. The rest part, $\tilde L_{2}$, of
the LD (\ref{e2}) contains the terms
\begin{eqnarray}
\label{e23} \tilde L_{2} = e\,\alpha\, m\,A^{2}_{\mu}\,\psi +
e\,\alpha^{2}\, A^{\mu} (\psi\partial_{\mu}\chi -
\chi\partial_{\mu}\psi) +
\frac{1}{2}\,e^{2}\,\alpha^{2}\,A^{2}_{\mu} (\psi^{2} + \chi^{2})
\cr -\frac{1}{4}\,\lambda^{2}(\psi^{2} + \chi^{2})^{2} -
\lambda^{2}\,\sigma\,\psi(\psi^{2} + \chi^{2}) + \sigma
(\mu^{2}_{0}- \lambda^{2}\sigma^{2})\psi.
\end{eqnarray}
At the lowest order the condition $(\Omega,\psi\Omega) = 0$
imposes
$\mu^{2}_{0}= \lambda^{2}\sigma^{2}$.

Our aim is the canonical quantization of (\ref{e22}). The
equations of motion are
\begin{eqnarray}
\label{e24} (\alpha^{2}\,\nabla^{2} + \mu^{2})\psi = 0,
\end{eqnarray}
\begin{eqnarray}
\label{e25} \alpha\,\nabla^{2}\chi + m\,\partial_{\mu}A^{\mu} = 0,
\end{eqnarray}
\begin{eqnarray}
\label{e26} \nabla^{2} A_{\mu} - \beta_{U}\,
 \partial_{\mu}\partial^{\nu}A_{\nu} + m^{2}\,A_{\mu} +
\alpha\,m\,\partial_{\mu}\chi = 0,
\end{eqnarray}
where $\mu^{2} = 2\,\lambda^{2}\sigma^{2}$.
The pattern related to the field $\psi$ presents no problems and
will not be considered. By taking the divergence of equation
(\ref{e26}) and using the equation (\ref{e25}), one gets
\begin{eqnarray}
\label{e30} \nabla^{2}\partial_{\mu} A^{\mu} =0
\end{eqnarray}
from which the dipole-type equation, that resembles the Froissart
model [5], can be easily obtained
\begin{eqnarray}
\label{e31} \lim_{\kappa^{2}\rightarrow 0} (\nabla^{2} +
\kappa^{2})^{2}\,\chi =0.
\end{eqnarray}
This means that the scale-invariant particle staff is almost
massless. The "spectrum" of the model consists of unparticle stuff
and massive bosons. The solution of the Eq. (\ref{e26}) is found
in the form
\begin{eqnarray}
\label{e32}
A_{\mu}(x) = B_{\mu} (x) - \frac{\alpha}{m}\,\partial_{\mu}\left
[ 1 - \frac{1 - \beta_{U}}{m^{2}}\,\nabla^{2}\right ]\chi (x),
\end{eqnarray}
where the field $B_{\mu}$ obeys $(\nabla^{2} + m^{2})B_{\mu} = 0$,
$\partial_{\mu}B^{\mu} = 0$ and $[B_{\mu}(x), \chi (y)] =0$.
The massive vector field  $B_{\mu}(x)$ is derived from the vector
potential  $A_{\mu}(x)$ shifted by terms containing the
divergences of unHiggs field at finite $\sigma$.  The unparticle
sector no longer remains scale-invariant because of electroweak
symmetry breaking.

We find that under the gauge transformations (\ref{e17}),
(\ref{e18}) the generating current is defined by the vector field
$B_{\mu}(x)$ with the mass $m = e\,\alpha\,\sigma$ and the
unHiggs field $\chi (x)$
\begin{eqnarray}
\label{e40}
K_{\mu}(x)= - m^{2}\left [B_{\mu}(x) +
\frac{1-\beta_{U}}{m^{3}}\,\alpha\,\partial_{\mu}\nabla^{2}\chi(x)\right
],
\end{eqnarray}
where $\sigma$ is the  solution of eq. (\ref{e13}).

To find the propagator of $\chi(x)$ - field we use the two-point
Wightman functions (TPWF) formalism of a scalar field satisfying
the dipole-type equation.

\section{Quantization of the "unHiggs" field}

The field $\chi(x)$ obeying the equation
${(\nabla^{2})}^{2}\chi(x) = 0$ in four-dimension space-time can
be easily obtained as a result of quantization of the system of
two scalar neutral fields $\chi(x)$ and $\zeta (x)$ in the
quadratic LD [7]
\begin{eqnarray}
\label{e301}\tilde L = \partial_{\mu}\chi(x)\partial^{\mu}\zeta
(x) + \frac{1}{2}\zeta^{2}(x).
\end{eqnarray}
The equations of motion
\begin{eqnarray}
\label{e311}\nabla^{2}\chi (x) = \zeta (x)
\end{eqnarray}
and
\begin{eqnarray}
\label{e3111}\nabla^{2}\zeta (x) = 0
\end{eqnarray}
allow to exclude the field $\zeta (x)$ and get the dipole equation
(\ref{e31}) in the limit $\kappa =0$. The only equal-time
nontrivial canonical commutation relation (CCR)
\begin{eqnarray}
\label{e301} [\partial_{0}\zeta (x),\chi(y)]_{x^{0}=y^{0}} =
-i\,\delta (\vec x - \vec y) = [\partial_{0}\chi(x),\zeta
(y)]_{x^{0} = y^{0}}
\end{eqnarray}
together with equations of motion (\ref{e311}), (\ref{e3111})
leads to CCR for the unHiggs field $\chi(x)$
\begin{eqnarray}
\label{e33} [\chi(x),\chi(y)] = -i\,b_{d}\,E(x-y),
\end{eqnarray}
where
\begin{eqnarray}
\label{e302} E(x) = i\int 2\pi sign(p^{0})\delta^{\prime} (p^{2})
e^{-ipx}\frac{d^{4}p}{(2\pi)^{4}} =
\frac{1}{8\,\pi}\,sign(x^{0})\theta(x^{2}).
\end{eqnarray}
Here, $b_{d}$ (in mass dimension $2d$) ensures the correct mass
dimension of (\ref{e33}); $sign (x)$ and $\theta (x)$ are the sign
and the step functions, respectively.

The next step is to decompose $E(x)$ into its negative
($E^{-}(x)$) - and positive ($E^{+}(x) = [E^{-}(x)]^{*} = -
E^{-}(-x) $) -frequency parts, each of which is analytic in the
past and future tubes: $E(x) = E^{-}(x) + E^{+}(x)$ with
\begin{eqnarray}
\label{e303} E^{-}(x) = i\int 2\pi\theta (p^{0})\delta^{\prime}
(p^{2}) e^{-ipx}\frac{d^{4}p}{(2\pi)^{4}} =
-\frac{i}{(4\pi)^{2}}\,\ln \frac{l^{2}}{-x^{2} + i\epsilon x^{0}}
\cr = -\frac{i}{(4\pi)^{2}} [\ln\vert\kappa^{2} x^{2}\vert + i\pi
sign (x^{0})\theta (x^{2})].
\end{eqnarray}
Here, $l$ is an arbitrary length scale with dimension minus one in
mass units and introduced in the logarithmic function $\ln
[-(x^{0} - i\epsilon)^{2} + \vec x^{2} ]$ for dimensional reason
and $\kappa\sim l^{-1}$ being the mass parameter of the IR
regularization. The origin of $l$ becomes more transparent in
momentum space. Note that the distribution $\theta
(p^{0})\delta^{\prime} (p^{2})$ in (\ref{e303}) is well-defined
only with the basic functions $u(p)$ having the properties: $u(p)
= 0$ at $p = 0$.

The unHiggs field $\chi(x)$ is similarly decomposed into negative
(annihilation, $\chi^{-}(x)$) - and positive (creation,
$\chi^{+}(x) = [\chi^{-}(x)]^{+}$) - frequency parts $\chi(x) =
\chi^{-}(x)  + \chi^{+}(x)$ with commutators
\begin{eqnarray}
\label{e304} [\chi^{-}(x),\chi^{+}(y)] = - i\, b_{d}\,E^{-}(x-y),
\end{eqnarray}
\begin{eqnarray}
\label{e305} [\chi^{-}(x),\chi^{-}(y)] = [\chi^{+}(x),\chi^{+}(y)]
= 0.
\end{eqnarray}
It is assumed that the vacuum $\Omega =\vert 0\rangle $ satisfies
$\chi^{-}(x)\,\Omega = 0$ where $\langle \Omega,\Omega\rangle =
1$.

We introduce the TPWF
\begin{eqnarray}
\label{e34} \omega (x-y) = \langle 0\vert \chi(x)\chi(y)\vert
0\rangle
\end{eqnarray}
as the distribution in the Schwartz space $S^{\prime}(\Re^{4})$ of
the moderate distribution on $\Re^{4}$.
Taking into account the equation  (\ref{e31}) one has
\begin{eqnarray}
\label{e35} (\nabla^{2})^{2} \omega (x) = 0.
\end{eqnarray}
Because of the Lorentz invariance requirement the general solution
of eq. (\ref{e35}) is given in the form
\begin{eqnarray}
\label{e36} \omega (x) = b_{1}\,E^{-} (x) + b_{2}\,\nabla^{2}
E^{-} (x) + const,
\end{eqnarray}
where
\begin{eqnarray}
\label{e306} \nabla^{2} E^{-} (x) =
\frac{-i}{4\,\pi^{2}}\frac{1}{x^{2} - i\,\epsilon\,x^{0}},\,\,\,
(\nabla^{2})^{2}E^{-} (x) = 0,
\end{eqnarray}
$b_{1}$ and $b_{2}$  are coefficients. Therefore
\begin{eqnarray}
\label{e3006} [\chi (x),\chi (0)] = 2\,\pi\,i\,sign (x^{0})
[b_{1}\,\theta (x^{2}) + b_{2}\,\delta (x^{2})].
\end{eqnarray}

\section{Propagator of the "unHiggs" field}

The propagator of $\chi(x)$ is defined as the time-ordered TPWF
\begin{eqnarray}
\label{e40} W(x) = \langle 0\vert T\chi(x)\chi(0)\vert 0\rangle =
\theta (x^{0})\omega (x) + \theta (-x^{0})\omega (-x) = \cr
b_{1}ln\frac{l^{2}}{- x^{2} + i\epsilon} + b_{2} \left
[\frac{1}{x^{2}} + i\pi\delta (x^{2})\right ] + const.
\end{eqnarray}
For long range forces mediated by unparticles one has for
$W(x)$:
\begin{eqnarray}
\label{e401} W(x) \simeq b\,\ln\frac{l^{2}}{-x^{2} + i\epsilon} + const.
\end{eqnarray}

The coefficient $b$ in (\ref{e401}) can be fixed using the CCR
$[\alpha\chi(x), \pi(\chi(0))]_{x^{0}=0} = i\delta^{3}(\vec x)$
with $\pi(\chi(x)) = \alpha\partial_{0}\chi(x) + m\,A_{0} (x)$
being the conjugate momentum of $\chi (x)$.
Taking into
account the following properties of the generalized functions
\begin{eqnarray}
\label{e38} \nabla^{2} sign(x^{0})\,\theta (x^{2})=
4\,sign(x^{0})\,\delta(x^{2}),
\end{eqnarray}
\begin{eqnarray}
\label{e39}\frac{\partial}{\partial x^{0}}\left
[\frac{1}{2\,\pi}\, sign (x^{0})\,\delta (x^{2})\right ]_{x^{0} =
0} = \delta^{3}(\vec x),
\end{eqnarray}
we get the coefficient $b$ in the form:
\begin{eqnarray}
\label{e40} b
=\frac{(e\,\sigma)^{2}}{(4\pi)^{2}\,(1-\beta_{U})}.
\end{eqnarray}
Finally, we have

\begin{eqnarray}
\label{e41} W(x) \simeq \frac{(e\,\sigma)^{2}}{(1-\beta_{U})}
\bar W (x)+ const = \frac{(e\,\sigma)^{2}}{ (1-\beta_{U})}
\frac{1}{(4\pi)^2}\ln\frac{l^{2}}{-x^{2} + i\epsilon} + const,
\end{eqnarray}
where $\bar W(x)$ obeys the equation
\begin{eqnarray}
\label{e42} \left (\nabla^{2}\right )^{2} \bar W (x) =
-i\delta^{4} (x).
\end{eqnarray}

Actually, at energy $E < \tilde\Lambda_{U}$ parameter $\beta_{U}$
is equal to one, while $\sigma = 0$ because the interaction between
the Higgs boson and unparticle
disappears ($a=0$) in the continuum limit (see formulas (\ref{e3})
and (\ref{e13})). The presence of the parameter $l$ in TPWF
(\ref{e36}) (with $E^{-} (x)$ defined in (\ref{e303})) breaks its
covariance under dilatation transformations
$x\rightarrow\varpi\,x$ within which $\omega (x)\rightarrow \omega
(\varpi x) = \omega (x)- (8\pi^{2})^{-1}\,\ln \varpi$ for any
$\varpi >0$. This implies spontaneously symmetry breaking of the
dilatation invariance. This is one of the reasons for the special
role of the function $E^{-}(x)$ considered above.

Let us consider the v.e.v. $\langle 0\vert [K_{\mu}(x), \chi
(0)]\vert 0 \rangle$ in the framework of the Goldstone theory. The
equal-time commutator $[K_{0}(x),\chi (0)]$ integrated over $\vec
x$-space gives the v.e.v. of the unHiggs
\begin{eqnarray}
\label{e402} \sigma = \frac{i}{e}\,\int d^{3} x\,[K_{0}(x),\chi
(0)]_{\vert_{x^{0} =0}}
\end{eqnarray}
and we find that
\begin{eqnarray}
\label{e4002} i\,\langle 0\vert [K_{\mu}(x),\chi (0)]\vert
0\rangle = \frac{e\,\sigma}{2\,\pi}\,\partial_{\mu} \left [sign
(x^{0})\,\delta (x^{2})\right ].
\end{eqnarray}
The Fourier transformation of the r.h.s. of (\ref{e4002}) ensures
the term proportional to $p_{\mu}\,sign (p^{0})\,\delta (p^{2})$
where $\delta (p^{2})$ is a consequence of the Goldstone theorem.

The propagator $\tilde W(p)$ of the unHiggs field is introduced
according to
\begin{eqnarray}
\label{e4401} W(x) = \int d_{4} p\, e^{-i\,p\,x}\,\tilde W(p).
\end{eqnarray}
Because $W(x)$ satisfies $(\nabla^{2})^{2}\,i\,W(x) =
\varepsilon\,\delta^{4}(x)$, where $\varepsilon =
(e\,\sigma)^{2}\,(1-\beta_{U})^{-1}$ and $\nabla^{2}\,W(x) =
(\varepsilon /4\,\pi^{2})(-x^{2} + i\,\epsilon)^{-1}$, the
propagator $\tilde W(p)$ is a solution of the equation
\begin{eqnarray}
\label{e4402} p^{2}\, \tilde W(p) = \frac{i\,\varepsilon}{-p^{2} -
i\,\epsilon}
\end{eqnarray}
and furthermore one has
\begin{eqnarray}
\label{e4403} i\, \tilde W(p) = \frac{\varepsilon}{(-p^{2} -
i\,\epsilon)^{2}}
\end{eqnarray}
in the sence of deneralized functions (distributions), when the l.h.s. of
(\ref{e4402}) and (\ref{e4403}) act on functions which are regular at $p=0$.

Formally, the latter expression leads to IR divergence at small
$p^{2}$. However, one can image $W(x)$ in terms of generalized
functions, and its Fourier transform is guaranteed to exist.
Following [8] we introduce
\begin{eqnarray}
\label{e4404}  \tilde W(p) = \frac{\partial^{2}}{\partial
p^{2}}\,H(p),
\end{eqnarray}
where
\begin{eqnarray}
\label{e4405}  H(p) = \frac{\varepsilon}{16\,\pi^{2}}\,\int
d^{4}x\,e^{i\,p\,x}\,\frac{\ln\left (\frac {l^{2}}{-x^{2} +
i\,\epsilon}\right )}{-x^{2} + i\,\epsilon}.
\end{eqnarray}
The desired propagator $\tilde W(p)$ is given either in the form
\begin{eqnarray}
\label{e4406}  \tilde W(p) = - \frac{1}{2}\,i\,\varepsilon \,\frac
{\partial}{\partial p_{\mu}}\,\left [\frac{p^\mu\,\ln
(-\varsigma^{2}\,p^{2} - i\,\epsilon )}{(-p^{2} -
i\,\epsilon)^{2}}\right ]
\end{eqnarray}
or
\begin{eqnarray}
\label{e4407}  \tilde W(p) = -\frac{1}{4}\,i\varepsilon \,\frac
{\partial^{2}}{\partial p^{2}}\left \{\frac{\ln\left [e^{2\gamma}
(-p^{2}\,l^{2} - i\,\epsilon)/4\right ]}{-p^{2} -
i\,\epsilon}\right \} ,
\end{eqnarray}
where $\varsigma = e^{\gamma - 1/2}\, l/2$ is the regularized
length, $\gamma= - \Gamma^{\prime}(1) \simeq 0.577$.

To avoid the IR divergence the propagator $\tilde W(p)$
(\ref{e4406}) has to be considered as the generalized function
(distribution) with the basic (test) function $f(p)$
\begin{eqnarray}
\label{e4408} \int d_{4}p\, \tilde W(p)\,f(p) =
\frac{1}{2}\,i\,\varepsilon\, \int d_{4}p\, \frac{\ln
(-\varsigma^{2}\,p^{2} - i\,\epsilon )}{(-p^{2} -
i\,\epsilon)^{2}} p^\mu \frac {\partial}{\partial p_{\mu}}\,f(p).
\end{eqnarray}
The extra power of $p_{\mu}$-momentum in (\ref{e4408}) removes the
divergence at small $p$.
 On the other hand, in order to find the Fourier transformation of $W(x)$
 in  (\ref{e41}) let us consider the equation
\begin{eqnarray}
\label{e43} \lim_{\lambda^{2}\rightarrow 0}\int d_{4}p\,
e^{-i\,p\,x}\,\frac{1}{(p^{2} - \lambda^{2} +
i\epsilon)^{2}} = \frac{i}{8\pi^{2}}\,K_{0}\left
(\lambda\sqrt{-x^{2} + i\epsilon}\right).
\end{eqnarray}
Here, $K_{0}(z)$ is the Bessel function; $\lambda$ is the
parameter of representation and not the analogue of the infra-red
mass. The naive use of (\ref{e36}) leads to appearance of
$\ln\lambda$ divergence, because at small $z$ the following
behaviour of $K_{0}(z)$
\begin{eqnarray}
\label{e44} \lim_{z\rightarrow 0} K_{0}(z) \simeq  \ln (2/z)
-\gamma + O(z^{2}, z^{2}\ln z)
\end{eqnarray}
is correct.
To avoid the $\ln\lambda$ divergence one can use the following
distribution for the Fourier transformation of $W (x)$
\begin{eqnarray}
\label{e45} \lim_{\lambda^{2}\rightarrow 0}\int d_{4}p\,
e^{-i\,p\,x}\,\left [\frac{1}{(p^{2} - \lambda^{2}
+ i\epsilon)^{2}} +
\frac{i}{(4\pi)^{2}}\,\ln\frac{\lambda^{2}}{\kappa^{2}}\,
\delta^{4}(p)\right ]
\end{eqnarray}
with the mass parameter $\kappa\sim l^{-1}$. The $const$ term in
eq. (\ref{e41}) can be fixed in such a way as to cancel the
constant terms  occurred due to expansion (\ref{e44}). Thus the
propagator for the unHiggs field $\chi(x)$ becomes in momentum $p$
-  space
\begin{eqnarray}
\label{e46} \tilde W (p) = \frac{(e\,\sigma)^{2}} {(1-\beta_{U})}
\lim_{\lambda^{2}\rightarrow 0}
\left [\frac{1}{i\,(p^{2} - \lambda^{2} + i\epsilon)^{2}} +
\frac{1}{(4\pi)^{2}}\,\ln\frac{\lambda^{2}}{\kappa^{2}}\,
\delta^{4} (p)\right].
\end{eqnarray}
The distribution $\lim_{\lambda^{2}\rightarrow 0} [1/(p^{2} -
\lambda^{2} + i\epsilon)^{2}]$ is defined only on a particular
subspace of the space of complex Schwartz test functions on $\Re
^{4}$, namely on those (test) functions $f(p)$ such that $f(0)
=0$. The set of its extensions onto the whole space is a
one-parameter set of functionals parameterized by $\kappa$. It means that
the distribution $(p^{2} + i\epsilon)^{-2}$ is defined only on
functions vanishing at $p = 0$. However, the set of Lorentz-invariant,
causal extensions of this distrubution to those not vanishing at $p = 0$
constitute a new parameter ($\kappa$) family given by (\ref{e45}).

A consequence of $\tilde W(p)$ is that the lowest order energy
(potential) of a "charge" in the static limit has the following
Fourier transform $(\vert \vec x \vert \equiv r)$
\begin{eqnarray}
\label{e47} E(r;\sigma) = i\,\int d_{3}\vec {p}\,e^{i\,\vec p\,\vec
x}\, \tilde W (p^{0} =0, \vec p; \sigma)=\cr
\frac{(e\,\sigma)^{2}}{8\,\pi\,(1-\beta_{U})}\,r\left
[\frac{9}{2}\,\ln 2 -2 + 3\,\ln(\kappa\,r)\right ]
\end{eqnarray}
by making use the representation (\ref{e4407}) in the static limit
\begin{eqnarray}
\label{e48} i\,\tilde W (p^{0} = 0, \vec p; \sigma) =
\frac{(e\,\sigma)^{2}}{2\,(1-\beta_{U})}\frac{1}{({\vec p^{2}}
-i\epsilon)^{2}}\cr \times\left [\left (1-\frac{4\,\vec
p^{2}}{\vec p^{2} -i\epsilon}\right )\ln\left (\frac{\vec
p^{2}\,l^{2} -i\epsilon}{4}\,e^{2\gamma}\right ) + \frac{6\, \vec
p^{2}}{\vec p^{2} -i\epsilon} -1 \right ].
\end{eqnarray}
The energy (\ref{e47}) grows as $r$ at large distances. To get the
result (\ref{e47}) we used the formula [9] ($p\equiv \vert \vec
p\vert$)
\begin{eqnarray}
\label{e49} \int d^{3}p e^{i\vec p\vec x} p^{n} \ln^{m} p =
r^{-(n+3)}\sum_{i=0}^{m}\frac{(-1)^{i}\Gamma (m+1)}{\Gamma
(m-i+1)\Gamma (i+1)}\left (\frac{\partial}{\partial n}\right )
^{m-i}C_{n}\ln^{i} r,
\end{eqnarray}
where \begin{eqnarray}\label{e49} C_{n} =
2^{n+3}\,\pi^{3/2}\,\frac{\Gamma [(n+3)/2]}{\Gamma (-n/2)}, \,\,\,
n\neq -3, -5, ...
\end{eqnarray}


\section{Conclusion}
\label{bec}
We have presented the gauge model of scalar unparticles.
From the phenomenological point of view the most physical
motivated way to explore the unparticle sector is the interaction
with the Higgs field operator ${\vert H\vert}^{2}$. We have
clarified how the scale-invariant self-coupling $\xi$ generates
the mass squared $m^{2}_{IRR} = b^{\prime}\,v^{2} -
\xi\,\sigma^{2}$ for unparticles that acts
 as an infra-red cutoff to give a nonzero $\sigma = \langle O\rangle$.

Because we have the exact solution for the unparticle TPWF, one
can see precisely how the system makes the transition from high
energy physics of free particles to low energy unparticle physics.

The scalar unHiggs model has been studied at the lowest order
of perturbative series, using the canonical quantization.
The new dipole-type behaviour of the propagator for the unHiggs field
has been found (ghost-like field). Moreover the vector field becomes massive.
In particular, the unHiggs field $\chi (x)$ united with $A_{\mu} (x)$- field
both form the field $B_{\mu} (x)$ in such a way that the latter obey
the equation
of motion for massive vector field with the mass $m =\alpha\,e\,\sigma$.
The Goldstone theorem is verified as the statement of the vacuum
expectation value of the commutator of generating current $K_{\mu}(x)$ and of
(appropriate) local field - the unHiggs field $\chi (x)$.

The following observations concerning the unHiggs propagator are in order:\\
1. $\tilde W(p)$ is valid in the energy region $\tilde\Lambda_{U}
< E < \Lambda_{U}$;\\
2. the form (\ref{e46}) is uniquely determined by the canonical formalism;\\
3. it clearly does not violate causality because of (\ref{e302}).

Formula (\ref{e47}) explicitly shows how large logarithms of
dynamical unHiggs propagator origin are suppressed at short
distances. The behaviour of (\ref{e47}) is the most general and
compelling argument in favor of the (free) energy of unHiggs
field.

\section{Acknowledgments}

It is a pleasure to thank Laboratories which I visited during the
course of this work: CERN and Fermilab. I am grateful for
stimulating conversations with colleagues, particularly N.
Krasnikov.


\end{document}